\documentclass[a4paper,twoside,10pt]{article}

\usepackage[T1]{fontenc}
\usepackage[ansinew]{inputenc}
\usepackage{lmodern}
\usepackage{xcolor}

\usepackage{amsmath}
\usepackage{amsthm}
\usepackage{amsfonts}

\usepackage{graphicx} 
\usepackage[section]{placeins}
\usepackage{enumerate}

\usepackage[style=authoryear, defernumbers=true, backend=biber,dashed=false, maxnames=999,maxcitenames=3,giveninits=true,urldate=long,uniquename=false,uniquelist=false
]{biblatex}

\begin{document}

\pagestyle{empty} 

\pagestyle{plain} 


    \vspace*{\stretch{1.0}}
   \begin{center}
      \Large\textbf{The performance of the partially overlapping samples t-tests at the limits.}\\
      \large\textit{B. Derrick, D. Toher, P. White}
    \vspace*{\stretch{1.0}}
   \end{center}

\begin{abstract}
    This note includes an assessment of the partially overlapping samples t-tests in scenarios where elements of the test are at their extremes, including where only one sample contains independent observations
\end{abstract}
 
     \vspace*{\stretch{1.0}}

\section{Background}
Tests for the comparison of means for two groups that contain both independent and paired observations are proposed by Derrick et al. (2017). Their simulations show this solution is robust for a wide variety of scenarios. In this note, the performance of these tests is considered when only one of the two groups have independent observations, and in other extreme parameter conditions.

The simulation methodology outlined in Derrick et al. (2017), is used to assess the Type I error robustness and power of the authors partially overlapping samples t-tests; namely $T_{new1}$ and $T_{new2}$. These are compared against the paired samples t-test discarding independent observations, $T_1$, the independent samples t-test discarding pairs, $T_2$, and Welch's test discarding pairs, $T_3$.

A factorial design was considered with 10,000 iterations of each parameter combination, with a missing completely at random (MCAR) mechanism. The parameters used within the original simulation deign are given in Table \ref{ta:simpars}. 

\begin{table}[!htb]
\caption{Simulation parameters}\label{ta:simpars}
\centering
\begin{tabular}{ll}
\hline\hline
\textbf{Parameter}	&	\textbf{Values}	\\
\hline
$\mu_{1}$& 0  \\
$\mu_{2}$& 0 (under $H_{0}$); 0.25, 0.50, 0.75, 1.00, 1.25, 1.50 (under $H_{1}$) \\
$\sigma_{1}^{2}$& 1, 2, 4, 8 \\
$\sigma_{2}^{2}$& 1, 2, 4, 8\\
$n_{a}$& 5, 10, 30, 50, 100, 500 \\
$n_{b}$& 5, 10, 30, 50, 100, 500 \\
$n_{c}$& 5, 10, 30, 50, 100, 500 \\
$\rho$& 0.75, -0.50, -0.25, 0.00, 0.25, 0.50, 0.75 \\
\hline\hline
\end{tabular}
\end{table}

In the notation, $n_a$ represents the number of independent observations in Sample 1, $n_b$ represents the number of independent observations in Sample 2, and $n_c$ represents the number of pairs.

\section{Extreme scenarios}

The following explores whether the partially overlapping samples t-tests remain valid for circumstances which include parameters of the test at their limits. Particular attention is given to the validity of the `Partover.test' function in the R package `partiallyoverlapping' by Derrick (2017). Simulations are performed to assess the Type I error robustness for the following extreme or unusual conditions:

\begin{enumerate}

\item $n_{a}$ = 0 and $n_{b}$ = 0

\item $n_{a}$ = 0 or $n_{b}$ = 0

\item $n_{c}$ = 0 or $n_{c}$ = 1

\item $\rho$ = 1 or $\rho$ = -1 and for the paired observations the mean difference $d \neq$ 0

\item $\rho$ = 1 and for the paired observations the mean difference $d$ = 0

\item $\sigma^2$ = 0

\item $H_0:\mu_1 - \mu_2 = x $

\end{enumerate}

     \vspace*{\stretch{1.0}}

\subsubsection*{$n_{a} = 0$ and $n_{b} = 0$}

If there are no independent observations, both forms of the partially overlapping samples t-test are equivalent to the paired samples t-test. The `Partover.test' function in R can be performed and gives equivalent results to the paired samples t-test. 

\subsubsection*{$n_{a} = 0$ or $n_{b} = 0$}

If paired observations are present and only one sample has independent observations, incompleteness is in a single response only. The partially overlapping samples t-tests have the advantage that this is not a restriction to performing the `Partover.test'. See Table \ref{ta:limits1} for assessment of Type I error robustness for a selection of parameter combinations under these conditions.

\subsubsection*{$n_{c} = 0$ or $n_{c} = 1$}

If there are no paired samples, $T_{new1}$ is mathematically equivalent to the independent samples t-test, and is mathematically equivalent to Welch's test. When using the `Partover.test' in R, a minimum of $n_{c}$=2 is required due to the required calculation of a correlation coefficient. If $n_{c}$ =1, the pair must be discarded, the independent samples t-test or Welch's test could be performed without greatly impacting power.

\subsubsection*{$\rho = 1$ or $\rho = -1$ and for the paired observations $d \neq 0$}

In the extremes, it is theoretically possible to have perfect correlation, without the paired observations being identical. 

The simulation design is extended to consider the conditions above. Results from this extension to the simulation design for selected parameter combinations are given in Table \ref{ta:limits1}. Parameter combinations which fulfill  Bradley's liberal Type I error robustness are highlighted in bold.

\subsubsection*{$\rho = 1$ and for the paired observations $d = 0$}

There may be occasions by design where two subsets of the same group are compared, and some units are common to both subsets. For example, in education where the mean module score for `Statistical Modelling' is to be compared for two groups, those taking the optional module `Mathematical Statistics' and those taking the optional module `Operational Research'. Students taking both `Mathematical Statistics' and `Operational Research' could be said to be `paired' observations. For the students taking both optional modules, the score for the `Statistical Modelling' module has $\rho = 1$ and $d = 0$. The `paired' observations could be discarded and the independent samples t-test, $T_{2}$, or Welch's test, $T_{3}$, could be performed. Instead, the situation could be viewed as a one-way ANOVA, with three groups consisting of the two sets of independent observations, and the one set of paired observations. Alternatively the partially overlapping samples t-tests could be applied.

To assess whether the proposed test statistics are valid when $\rho = 1$ and for the paired observations $d = 0$, an additional consideration is required in the simulation design with respect to the variance. The variance of the paired observations, $\sigma_{c}^2$, could be equal to the variance of the observations in Group 1, $\sigma_{a}^2$, or the variance of the observations in Group 2, $\sigma_{b}^2$, or both, or neither. 

Table \ref{ta:limits2} give results of an extension to the simulation design to take into account these properties. Parameter combinations which fulfill Bradley's liberal Type I error robustness are highlighted in bold.

\begin{table}[!htb]
\caption{Type I error rates, extended design}\label{ta:limits1}
\centering
\begin{tabular}{cccccccc}
\hline\hline
$\rho$	&	$n_{a}$ &  $n_{b}$ & $n_{c}$ & $\sigma_{1}^{2}$ & $\sigma_{2}^{2}$ & $T_{new1}$ &  $T_{new2}$	\\
\hline

0.25	&	0	&	30	&	5	&	1	&	1	&	\textbf{ 	0.045	}	&	\textbf{ 	0.053	}	\\
0.75	&	0	&	5	&	5	&	1	&	1	&	\textbf{ 	0.041	}	&	\textbf{ 	0.040	}	\\
0.25	&	0	&	30	&	5	&	4	&	1	&		0.222		&	\textbf{ 	0.053	}	\\
0.75	&	0	&	5	&	5	&	4	&	1	&		0.124		&	\textbf{ 	0.048	}	\\
0.25	&	0	&	30	&	5	&	1	&	4	&		0.001		&	\textbf{ 	0.051	}	\\
0.75	&	0	&	5	&	5	&	1	&	4	&	\textbf{ 	0.032	}	&	\textbf{ 	0.041	}	\\
1	&	5	&	30	&	5	&	1	&	1	&	\textbf{ 	0.044	}	&	\textbf{ 	0.047	}	\\
1	&	30	&	5	&	5	&	1	&	1	&	\textbf{ 	0.042	}	&	\textbf{ 	0.047	}	\\
1	&	5	&	30	&	5	&	4	&	1	&	\textbf{ 	0.050	}	&	\textbf{ 	0.055	}	\\
1	&	30	&	5	&	5	&	4	&	1	&	\textbf{ 	0.044	}	&	\textbf{ 	0.050	}	\\
1	&	5	&	30	&	5	&	1	&	4	&		0.003		&	\textbf{ 	0.042	}	\\
1	&	30	&	5	&	5	&	1	&	4	&		0.185		&	\textbf{ 	0.050	}	\\
1	&	0	&	30	&	5	&	1	&	1	&	\textbf{ 	0.039	}	&	\textbf{ 	0.058	}	\\
1	&	0	&	5	&	5	&	1	&	1	&		0.019		&	\textbf{ 	0.044	}	\\
1	&	0	&	30	&	5	&	4	&	1	&	\textbf{ 	0.037	}	&	\textbf{ 	0.066	}	\\
1	&	0	&	5	&	5	&	4	&	1	&		0.019		&	\textbf{ 	0.043	}	\\
1	&	0	&	30	&	5	&	1	&	4	&		0.264		&	\textbf{ 	0.061	}	\\
1	&	0	&	5	&	5	&	1	&	4	&		0.180		&	\textbf{ 	0.064	}	\\
\hline\hline
\end{tabular}
\end{table}

\begin{table}[!htb]
\caption[Type I error rates, extended design]{Type I error rates, $n_{a}=5$ $\rho = 1$ and paired $d = 0$}\label{ta:limits2}
\centering
\begin{tabular}{cccccccccc}
\hline\hline
$n_{b}$ & $n_{c}$ & $\sigma_{a}^{2}$ & $\sigma_{b}^{2}$ & $\sigma_{c}^{2}$ & $T_{new1}$ &  $T_{new2}$ & $T_{1}$ & $T_{2}$ & ANOVA\\

	5	&	5	&	1	&	1	&	1	&	\textbf{ 	0.035	}	&	\textbf{ 	0.031	}	&	\textbf{ 	0.048	}	&	\textbf{ 	0.043	}	&	\textbf{ 	0.049	}	\\
	30	&	5	&	1	&	1	&	1	&	\textbf{ 	0.047	}	&	\textbf{ 	0.048	}	&	\textbf{ 	0.051	}	&	\textbf{ 	0.056	}	&	\textbf{ 	0.053	}	\\
	5	&	30	&	1	&	1	&	1	&	\textbf{ 	0.048	}	&	\textbf{ 	0.046	}	&	\textbf{ 	0.054	}	&	\textbf{ 	0.047	}	&	\textbf{ 	0.050	}	\\
	5	&	5	&	1	&	4	&	1	&		0.075		&	\textbf{ 	0.058	}	&	\textbf{ 	0.059	}	&	\textbf{ 	0.051	}	&	\textbf{ 	0.070	}	\\
	5	&	5	&	1	&	1	&	4	&		0.004		&		0.003		&	\textbf{ 	0.052	}	&	\textbf{ 	0.045	}	&	\textbf{ 	0.042	}	\\
	5	&	5	&	1	&	4	&	4	&		0.021		&		0.015		&	\textbf{ 	0.055	}	&	\textbf{ 	0.046	}	&	\textbf{ 	0.057	}	\\
	30	&	5	&	1	&	4	&	1	&		0.009		&	\textbf{ 	0.063	}	&		0.004		&	\textbf{ 	0.049	}	&		0.092		\\
	5	&	30	&	1	&	4	&	1	&		0.000		&		0.000		&	\textbf{ 	0.054	}	&	\textbf{ 	0.049	}	&	\textbf{ 	0.071	}	\\
\hline
\hline\hline
\end{tabular}
\end{table}

It can be seen from Table \ref{ta:limits1} that when variances are equal, both partially overlapping samples t-tests remain valid when one sample has no independent observations. Additionally when relaxing the equal variances assumption, $T_{new2}$ remains valid when the incompleteness is in a single sample. Perfectly correlated data does not detract from the validity of the tests. 

Table \ref{ta:limits2} shows that $T_{new1}$ and $T_{new2}$ are only valid if the variance in the paired observations is equal to the variance of both sets of independent observations. In general terms if the variance of the paired observations is not equal to the variance of the independent observations, it is likely that the observations are actually from separate populations, therefore a one way ANOVA may be more appropriate. In the circumstances summarised above where the partially overlapping samples t-test may be valid, the exact form of the research question is taken into consideration when selecting the appropriate test.

\subsubsection*{$\sigma^2 = 0$}

If the variability of the differences in the paired observations is equal to zero, the paired samples t-test cannot be performed. If there is no variability in the differences in the independent observations, the independent samples t-test cannot be performed. The partially overlapping samples t-tests remain functional in both these instances, so long as there is variability in either the independent observations or the paired observations.

It is possible that the paired observations within one sample could be constant (particularly for discrete data). This results in zero variability within the paired sample. Where the paired samples t-test in R reports an error, the `Partover.test' function is set to proceed with $r = 0$. This is a valid approach because the partially overlapping samples t-test has been shown to be valid for $\rho = 0$. In this scenario the partially overlapping samples t-statistic is identical to the independent samples t-statistic performed on all of the available data, however the degrees of freedom differ because the partially overlapping samples t-tests incorporate the size of the paired sample. This means that the degrees of freedom are lower for the partially overlapping samples t-test than the independent samples t-test on all of the data. The partially overlapping samples t-test in this extreme scenario is therefore less powerful than performing the independent samples t-test on all of the available data. 

In the event of zero variability within both samples, `Partover.test' is set to give a p-value of 1 if $\bar{x}_{1}$ - $\bar{x}_{2}$  = 0, else `Partover.test' gives a p-value of 0.

\subsubsection*{$H_0:\mu_1 - \mu_2 = x $}

The above simulations are concerned with testing whether there is a difference between two groups, which is the same as testing if the difference between the two groups is zero. There may be occasions where researchers wish to test whether the difference between the two groups is equal to some other fixed value. Cao et al. (2018) state Welch's test with the hypothesised difference in population means on the numerator. 

This extension to the numerator could be generalised to all forms of the t-test including the proposed partially overlapping samples t-test. Thus each t-test has the form:
\begin{equation*}
T=\frac{\bar{X_1}-\bar{X_2}-(\mu_1-\mu_2)}{stderr(\bar{X_1}-\bar{X_2})}
\end{equation*}

To demonstrate robustness of the t-tests when assessing against a null hypothesis of a defined difference $x$ between the two populations, the simulation design in Table \ref{ta:simpars} is repeated for $H_0: \mu_1-\mu_2=10$. 

The independent variates are generated using the `rnorm' function in R with Group 1 $\mu=10$ and Group 2 $\mu=0$. Correlated variates are calculated with $\mu=0$, and then $x=10$ is added each correlated variate in Group 1. Type I error rates for selected parameter combinations, and averaged across the entire simulation design are given in Table \ref{ta:nonzeromu}.

\begin{table}[!htb]
\caption[Type I error rates, $H_0:\mu_1-\mu_2=10$]{Type I error rates, $H_0: \mu_1-\mu_2=10$, $\sigma_{2}^{2}=1$,$n_{c}=5$ } \label{ta:nonzeromu}
\centering
\begin{tabular}{cccccccccc}
\hline\hline
$\rho$	&	$n_{a}$ &  $n_{b}$ &  $\sigma_{1}^{2}$ & $T_1$ & $T_2$ & $T_3$ & $T_{new1}$ & $T_{new2}$	\\
\hline
-0.75	&	5	&	10	&		1		&	\textbf{	0.051	}	&	\textbf{	0.050	}	&	\textbf{	0.051	}	&	\textbf{	0.051	}	&	\textbf{	0.051	}	\\
-0.50	&	10	&	30	&		4	&		\textbf{	0.051	}	&		0.160		&	\textbf{	0.053	}	&		0.129		&	\textbf{	0.048	}	\\
-0.25	&	30	&	5	&		1	&		\textbf{	0.048	}	&	\textbf{	0.052	}	&	\textbf{	0.059	}	&	\textbf{	0.051	}	&	\textbf{	0.052	}	\\
0.00	&	5	&	5	&		4	&		\textbf{	0.050	}	&	\textbf{	0.061	}	&	\textbf{	0.052	}	&	\textbf{	0.055	}	&	\textbf{	0.046	}	\\
0.25	&	10	&	5	&		1	&		\textbf{	0.049	}	&	\textbf{	0.055	}	&	\textbf{	0.054	}	&	\textbf{	0.047	}	&	\textbf{	0.046	}	\\
0.50	&	30	&	10	&		4	&		\textbf{	0.052	}	&		0.009		&	\textbf{	0.047	}	&		0.013		&	\textbf{	0.048	}	\\
0.75	&	5	&	30	&		1	&		\textbf{	0.051	}	&	\textbf{	0.047	}	&	\textbf{	0.054	}	&	\textbf{	0.042	}	&	\textbf{	0.043	}	\\
Overall & & & &  												\textbf{	0.050	}	&		0.101		&	\textbf{	0.051	}	&		0.079		&	\textbf{	0.049	}	\\
\hline\hline
\end{tabular}
\end{table}

For the cross-section of parameter combinations with equal variances in Table \ref{ta:nonzeromu}, $T_1$, $T_2$, $T_3$, $T_{new1}$ and $T_{new2}$ all demonstrate Type I error robustness for testing $H_0: \mu_1-\mu_2=10$. The Type I error rates for both equal and unequal variances follow the same pattern as when testing for $H_0: \mu_1-\mu_2=0$ thus the conclusions regarding robustness for each test generalise to any $H_0: \mu_1-\mu_2=x$

When performing the `Partover.test', assessment against a $H_0: \mu_1-\mu_2=x$ is performed by adding the command `mu = x'. 

\section{Summary}\

The statistic $T_{new2}$ is Type I error robust across all conditions simulated under normality and MCAR. The greater power observed for $T_{new1}$, compared to $T_{new2}$, under equal variances, is likely to be of negligible consequence in a practical environment. This is in line with empirical evidence for the performance of Welch's test, when only independent samples are present, which leads to many observers recommending the routine use of Welch's test under normality, e.g. Ruxton (2006).

For equal variances, $T_{new1}$ and $T_{new2}$ are Type I error robust. In addition they are more powerful than the standard Type I error robust approaches. When variances are equal, there is a slight power advantage of using $T_{new1}$ over $T_{new2}$, particularly when sample sizes are not equal. Under unequal variances, $T_{new2}$ is the most powerful Type I error robust statistic considered. 

When faced with a research problem involving two partially overlapping samples, if MCAR and normality can be reasonably assumed, the statistic $T_{new1}$ can be used when variances are equal. Otherwise under the same conditions when equal variances cannot be assumed the statistic $T_{new2}$ can be used. 

The proposed test statistics for partially overlapping samples provide a competitive alternative method for analysis with normally distributed data, that remain valid for extreme parameter values.

\section{References}
Cao, C., M. Pauly, and F. Konietschke (2018). The Behrens-Fisher Problem
with Covariates and Baseline Adjustments. In:
arXiv preprint arXiv:1808.08986

Derrick, B. (2017). Partiallyoverlapping: Partially overlapping samples t-tests.
In:CRAN

Derrick, B., B. Russ, D. Toher, and P. White (2017). Test statistics for the
comparison of means for two samples which include both paired observations
and independent observations. In:
Journal of Modern Applied Statistical
Methods 16.1, pp. 137-157.

Ruxton, G. D. (2006). The unequal variance t-test is an underused alternative
to Student's t-test and the Mann-Whitney U test. In: Behavioral Ecology 17.4 pp 688-690.

\end{document}